\documentclass[twocolumn]{openjournal}
\usepackage{graphicx}   % Including figure files
\usepackage{amsmath}    % Advanced maths commands
\usepackage{amssymb,amsfonts}    % Extra maths symbols

\shorttitle{L. A. Balona}
\shortauthors{L. A. Balona}

\begin{document}

\title{The extraordinary frequency pattern variation in $\delta$~Scuti stars}

\email{email: lab@saao.ac.za}

\author{L. A. Balona}

\affiliation{South African Astronomical Observatory, P.O. Box 9, Observatory
7935, Cape Town, South Africa}

\begin{abstract}
Inspection of the periodograms of {\em TESS} $\delta$~Scuti stars indicates 
that there is little, if any, similarity between the frequencies of stars
in the same region of the H--R diagram.  This is difficult to understand 
because pulsation models predict that stars with similar physical parameters 
should have similar frequencies.  To investigate the problem, a quantitative measure of similarity between frequency
patterns is described.  When applied to non-adiabatic pulsation models with 
similar temperatures and luminosities, a strong correlation is found between 
the frequency patterns, as expected.  The correlation remains high when rotational
frequency splitting is included.  When applied to observations, very little 
correlation can be found, confirming the impression from visual inspection.  It 
seems that each star has its own unique frequency pattern, unrelated to its 
position in the H--R diagram.  This presents a problem in our understanding of 
stellar pulsation.  The existence of a period-luminosity law and the effect of 
combination frequencies in $\delta$~Scuti stars is briefly discussed.
\end{abstract}


\section{Introduction}

Pulsational driving in models of A and F stars depends on the effective 
temperature, $T_{\rm eff}$, luminosity, $L/L_\odot$, and chemical composition.
Rotation has the effect of introducing additional frequencies. For stars 
located close together within a small region in the H--R diagram, the frequency 
spectra should be similar for stars with the same rotation rate.  Differences 
in chemical composition will also affect the frequencies, but to a lesser
extent.  Furthermore, all stars within the instability strip should pulsate.

Inspection of the periodograms of several thousand $\delta$~Sct stars observed 
by {\em TESS} gives a strong impression that the above conclusions are not 
correct.  \citet{Bedding2023} has recently remarked that the characteristics of
the pulsation spectra in {\em TESS} $\delta$~Sct stars are varied and do
not correlate with stellar temperature.  More than half of the stars within the
instability strip do not seem to pulsate at all \citep{Balona2018c}.  The 
frequency patterns in $\gamma$~Dor stars are very different from $\delta$~Sct 
stars, even though they might have the same effective temperature and 
luminosity.  This cannot be attributed to rotation because the distribution of 
projected rotational velocities for $\gamma$~Dor and $\delta$~Sct stars are
the same.

Pulsation models cannot account for these observations. A better understanding 
of why certain pulsation modes are selected in preference to others is
required.  The study of mode selection in pulsating stars involves 
non-adiabatic, non-linear calculations for non-radial modes in rotating
stars, which is beyond current capabilities. It is no surprise that scant 
attention has been payed to this problem.  The review by 
\citet{Dziembowski1993c} is still mostly relevant.  For more recent work, see 
the excellent review by \citet{Smolec2014}. 

The aim of this paper is to present a sample of periodograms to illustrate the 
problem.  A method which measures the difference in frequency patterns in a 
quantitative way, while allowing for effects of rotation and chemical 
abundance, is described.  This is applied to {\em TESS} $\delta$~Sct stars in 
several regions of the H--R diagram.  The size of each region is sufficiently 
small that stars within it may be considered to have the same effective 
temperature and luminosity.  

The calculations indicate that frequency patterns of stars within each
region do not match at all.  It is shown that rotation cannot account for
the disparity.  This means that there cannot be a general period - luminosity 
law for $\delta$~Sct stars \citep{Poretti2008, McNamara2011, Garg2010, 
Poleski2010, Cohen2012, Ziaali2019}. The idea that combination frequencies may 
contribute to the disparity in frequency spectra can also be discarded.

\section{Data and variability classification}

Each {\em TESS} sector consists of about 27\,d of continuous photometry with 
2-min cadence. While stars near the ecliptic equator are only observed for one 
sector, stars near the ecliptic poles are observed in every sector.  The light 
curves are corrected for instrumental signatures and long-term drift.  These 
are called pre-search data conditioning (PDC) light curves. The data are from 
sectors 1--66 of the {\em TESS} mission.  

As each sector becomes available,  the light curves and periodograms of 
stars with $T_{\rm eff} > 6000$\,K are used to classify them according 
to variability type.  The classification scheme follows that of the 
{\em General Catalogue of Variable Stars} (GCVS, \citealt{Samus2017}). In
this way, many thousands of stars have been classified. A full description of 
the identification and classification of {\em TESS} variables is given in 
\citet{Balona2022c}.  As expected, most of the variables are of the 
$\delta$~Sct type.  Among the 125000 stars already classified, there are 
14000 $\delta$~Sct stars.

The {\em TESS} data present several problems in the classification of light
variability.  It seems that distinct regions of instability, as predicted
by the models, do not actually exist.  There is no region in the H--R diagram 
entirely free of pulsating stars \citep{Balona2020a}.  In order to assign a 
meaningful variability class, arbitrary boundaries in effective temperature, 
$T_{\rm eff}$, and pulsation frequency, $\nu$, must be defined.  For example, 
the boundary between the $\delta$~Sct and Maia stars is set at $T_{\rm eff} 
= 10000$\,K.  The periodograms of these two classes of variable look the
same: the Maia stars seem to be just an extension of the $\delta$~Sct 
instability strip to the early B stars \citep{Balona2023a}.  The arbitrary 
boundaries in $T_{\rm eff}$ and $\nu$ are forced by the lack of a suitable 
theory, but are required to define the traditional classification scheme.

\section{Variety of frequency patterns in $\delta$~Sct stars}
  
\begin{figure}
\centering
\includegraphics[]{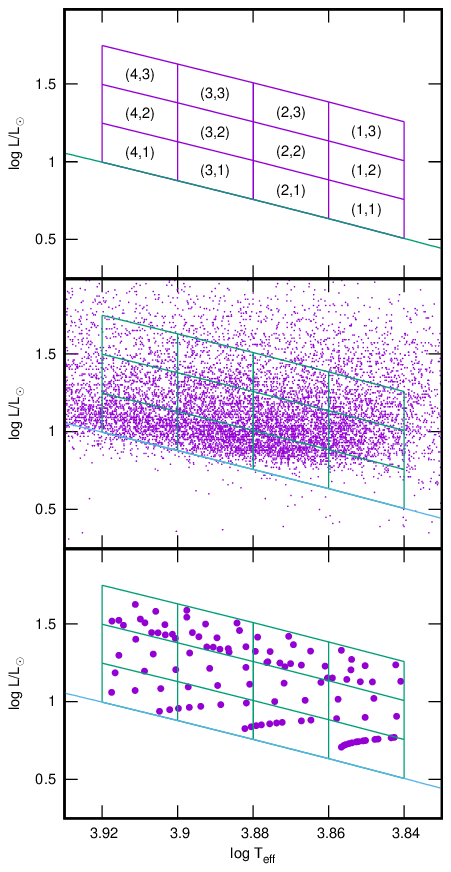}
\caption{The top panel shows the location of the regions in the H--R
diagram and the zero-age main sequence.  The middle panel shows $\delta$~Sct 
stars detected by {\em TESS}, while the bottom panel shows models used in 
the analysis.}
\label{reg}
\end{figure}

To illustrate the variety of frequency patterns in $\delta$~Sct stars with
similar stellar parameters, the instability strip was divided into 12
regions parallel to the zero-age main sequence (ZAMS), each region having a 
size $(\Delta \log T_{\rm eff}, \Delta \log L/L_\odot) = (0.02, 0.25)$.
This is sufficiently small that one may regard all stars within a region to
have approximately the same effective temperature and luminosity, 
while at the same time sufficiently large to include a suitable number of stars.
The region size also matches the uncertainties: $\sigma(\log T_{\rm eff}) \approx 
0.02$ and $\sigma (\log L/L_\odot ) \approx 0.1$.  Luminosities are from
{\em GAIA DR3} parallaxes \citep{Gaia2016, Gaia2021}.  Parameters of each 
region are listed in Table\,\ref{regions}. Fig.\,\ref{reg} shows the names and 
locations of the regions in the H--R diagram.

\begin{table}
\begin{center}
\caption{Data for rotating and non-rotating models located within a small
region in the H--R diagram. The first column is the region number followed
by the central $\log T_{\rm eff}$ and $\log L/L_\odot$ for the region. 
The last column is the number of models, $N_{\rm mod}$, or $N_{\rm obs}$,
the number of stars, within each region.}
\label{regions}
\begin{tabular}{crrrr}
\hline
\multicolumn{1}{c}{Region}                   & 
\multicolumn{1}{c}{$\log T_{\rm eff}$}       & 
\multicolumn{1}{c}{$\log\tfrac{L}{L_\odot}$} &
\multicolumn{1}{c}{$N_{\rm mod}$}            &
\multicolumn{1}{c}{$N_{\rm obs}$}           \\
\hline
(1,1) & 3.85  & 0.695 & 13 &  129  \\
(1,2) & 3.85  & 0.945 &  5 &  747  \\
(1,3) & 3.85  & 1.195 & 10 &  468  \\
 \\
(2,1) & 3.87  & 0.821 &  7 &  726  \\
(2,2) & 3.87  & 1.071 &  4 & 1048  \\
(2,3) & 3.87  & 1.321 & 14 &  459  \\
 \\
(3,1) & 3.89  & 0.943 &  8 & 1099  \\
(3,2) & 3.89  & 1.193 &  5 &  748  \\
(3,3) & 3.89  & 1.443 & 14 &  356  \\
 \\
(4,1) & 3.91  & 1.063 &  6 &  899  \\
(4,2) & 3.91  & 1.313 &  5 &  468  \\
(4,3) & 3.91  & 1.563 & 15 &  206  \\
\hline
\end{tabular}
\end{center}
\end{table}

Evolutionary stellar models were computed using the Warsaw - New Jersey
evolution code \citep{Paczynski1970}, assuming an initial hydrogen
fraction, $X_0 = 0.70$ and metal abundance, $Z = 0.020$ and using the chemical 
element mixture of \citet{Asplund2009} and OPAL opacities \citep{Rogers1992}.
Overshooting from the convective core was not included.   A mixing length
parameter $\alpha_{\rm MLT} = 1.0$ was adopted for the convective scale height. 
All models are non-rotating.  The non-adiabatic code developed by 
\citet{Dziembowski1977a} was used to obtain the pulsation frequencies and
growth rates.  Pulsations in these models are purely driven by the opacity 
$\kappa$~mechanism operating in the He\,II partial ionization region.  The
locations of the models in the H--R diagram are shown in the bottom panel of
Fig.\,\ref{reg}.

\begin{figure*}
\centering
\includegraphics[]{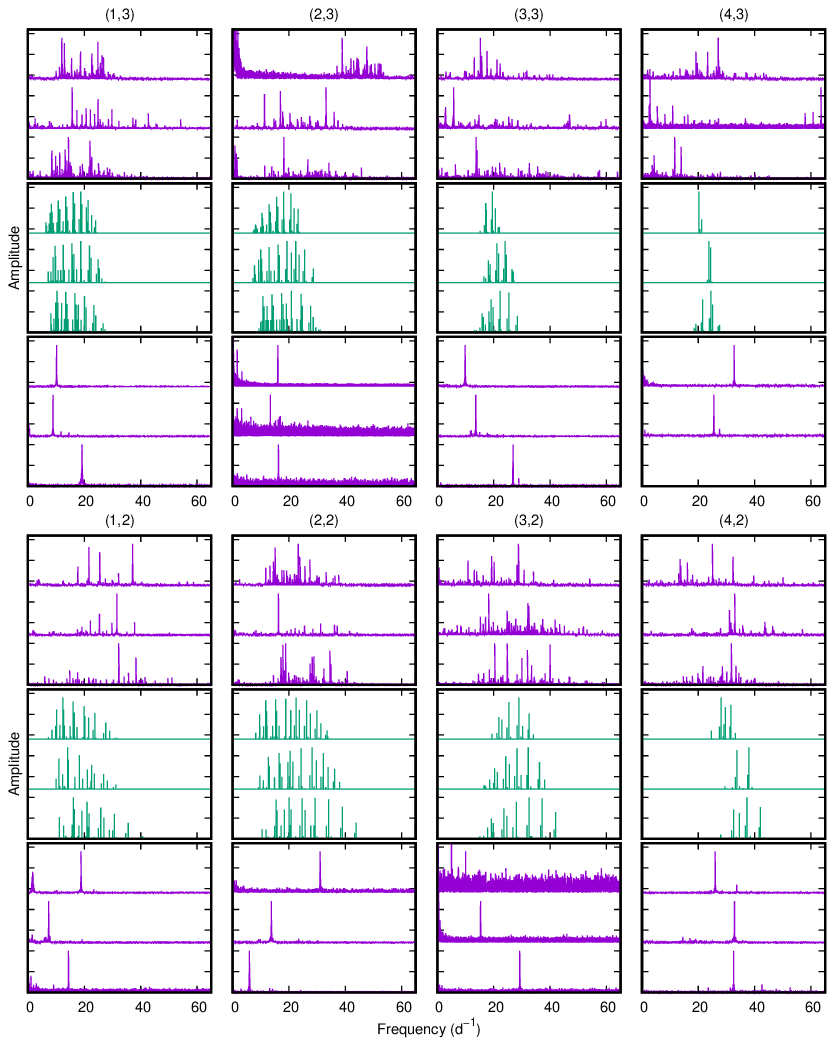}
\caption{Illustration of the variety of frequency patterns in $\delta$~Sct 
stars compared with theoretical models.  Each box consists of three panels.
The top panel shows stars with rich frequency spectra, while stars with just
a single dominant frequency peak are shown in the bottom panel.  The middle
panel (green) shows corresponding unstable frequencies from models.  The
label on the top of each box refers to a small region in the H--R diagram
(Table\,\ref{regions}).}
\label{freq}
\end{figure*}

\begin{figure*}
\centering
\includegraphics[]{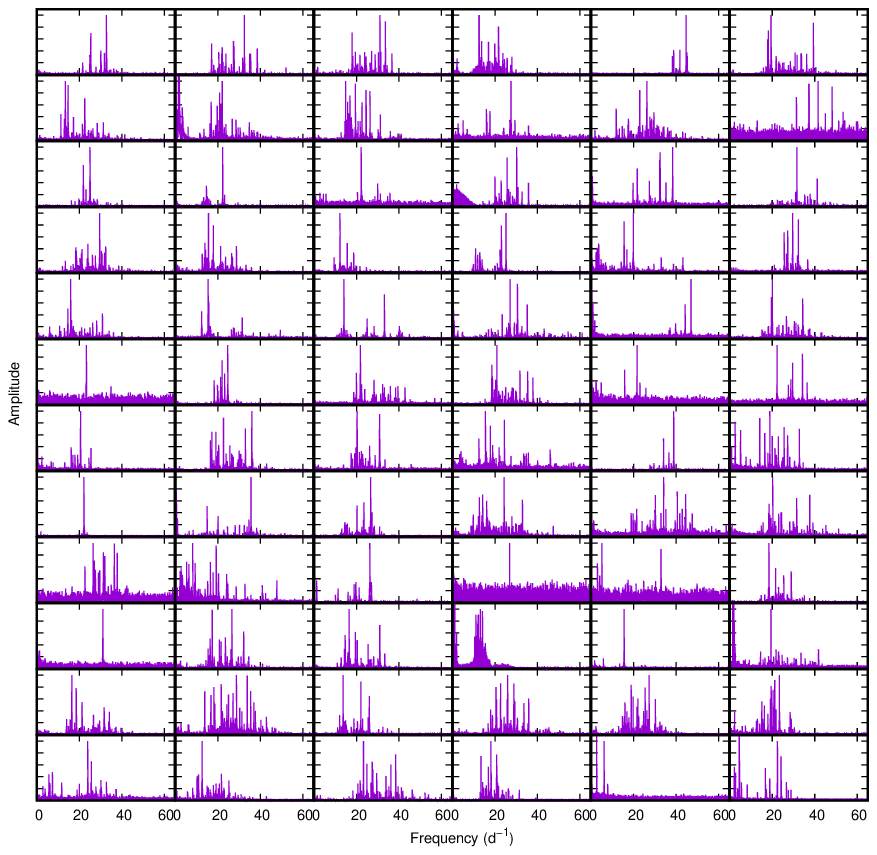}
\caption{Examples of {\em TESS} $\delta$~Scuti periodograms for stars within
a small region of the H--R diagram centered on $\log T_{\rm eff} = 3.87$,
$\log L/L_\odot = 1.07$ and with box size $0.02 \times 0.025$ dex.}
\label{reg22}
\end{figure*}

Examples of the disparity of frequency patterns within a given region are
illustrated in Fig.\,\ref{freq}.  In each region, examples of three stars 
with rich frequency spectra and three stars with just a single dominant 
frequency peak are shown.  In some cases one or two additional peaks are 
visible as well, but their amplitudes are always less than 5\% of the amplitude 
of the main peak.  About 5\% of $\delta$~Sct stars have a single dominant peak
defined in this way.

All stars have good spectroscopic estimates of $T_{\rm eff}$.  The middle row 
in Fig.\,\ref{freq} shows unstable modes from non-rotating models. The 
amplitudes were made to be proportional to the growth rate and corrected 
according to the visibility factor for the given spherical harmonic degree, 
$l$.  Modes with $l \le 8$ were selected. 

The figure shows that stars with the same physical parameters can have
completely different frequency patterns.  Frequencies of single-mode stars
may be very different from star to star within the same region. This indicates 
that the spherical harmonic degree and/or the radial order in these single-mode
stars must differ from star to star.  It is evident that the pattern in stars 
with rich spectra cannot be obtained from rotational frequency splitting of the 
single-mode pulsators.

While Fig.\ref{freq} shows stars across the H--R diagram, Fig.\,\ref{reg22} 
shows a larger sample of randomly chosen stars with similar, well determined, 
effective temperatures and luminosities in region (2,2). Each star seems to
have its own unique frequency pattern.

\section{Correlation}

While a subjective impression of the disparity of frequency patterns can be
obtained by inspection of Figs.\,\ref{freq} and \ref{reg22}, a method which
quantitatively measures the degree of similarity between a pair of frequency 
patterns is required.  One possibility is to calculate the correlation 
coefficient.

Suppose there are two frequency spectra, $i$ and $j$, each sampled with equal 
frequency spacing, $\delta \nu$, between frequencies $\nu_1$ and $\nu_2$.  Let 
$y_i(\nu)$ and $y_j(\nu)$ be the amplitudes at frequency $\nu$ of spectra $i$ 
and $j$.  The correlation coefficient, $r$, between the two spectra is given by
\begin{align*}
r = \frac{1}{A_iA_j} \sum_{\nu=\nu_1}^{\nu_2} y_i(\nu)y_j(\nu) \delta\nu,
\end{align*}
where $A_i$, $A_j$ are constants.  The values of these constants are chosen so 
that correlating a periodogram with itself gives unity.  Thus
\begin{align*}
A_i^2 = \sum_{\nu=\nu_1}^{\nu_2} y_i^2(\nu) \delta\nu,~~
A_j^2 = \sum_{\nu=\nu_1}^{\nu_2} y_j^2(\nu) \delta\nu.
\end{align*}
The value of $r$ lies between $r = 0$ (no similarity) and $r = 1$ 
(periodograms are identical).

The appearance of a frequency pattern depends not only on the frequencies,
but also on the signal-to-noise ratio, S/N, and the amplitudes.  Since all the 
observations discussed here were obtained by the same instrument, there should 
be no bias introduced by differing S/N ratios.  The S/N ratio is mostly 
determined by the brightness of a star and, to a lesser extent, by the length 
of the time series.  For stars within a given region, there will be 
approximately similar distributions of S/N and time series length, so these 
effects should average out. 

In calculating the correlation coefficient, $r$, each extracted frequency peak 
can either be given the same weight or weighted according to its amplitude.  
In the models, the amplitude of a given mode is taken to be proportional to the
growth rate.  A visibility factor which depends on the spherical harmonic, $l$,
is also applied.  Trials on models with unit weight lead to a slightly broader 
distribution of $r$, as might be expected.  The difference between weighting
the frequency according to amplitude or assigning unit weight to all observed 
peaks in $\delta$~Sct stars is almost non-existent.  This is owing to the
fact that $r$ is always close to zero, as demonstrated below.  In the results 
derived here, the frequency distributions are always weighted according to 
amplitude (with visibility factor applied for models), as this seems more 
realistic.  

Given any two stars, the extracted pulsation frequencies will never be exactly
the same. This means that the value of $r$ will always be near zero if the
frequency bin is very small. This problem is avoided by constructing a 
histogram of the frequency distribution with a suitable frequency bin size,
$\delta\nu$.  In selecting a suitable bin size, it is important to examine the 
range of frequencies that might be expected for stars within the same region
in the H--R diagram.

Examination of pulsation models within a range of 0.02\,dex in 
$\log T_{\rm eff}$ and 0.25\,dex in $\log L/L_\odot$ shows that for a fixed 
spherical harmonic degree and radial order $(l,n)$, the rms scatter in the 
frequencies is typically 2--6\,d$^{-1}$.  This can be higher for g modes or 
mixed modes.   A change in abundance leads to a somewhat different 
convective core size. Since g modes mostly sample the interior of a star near 
the convective core, there can be changes in frequencies of some g modes, but 
these are only of the order of several microhertz \citep{Guzik1998}.  This is 
smaller than the effect of rotation.  Other effects, such as mass loss and the 
presence of a magnetic field, do not alter the mean density and, in most
cases, have very little effect on the pulsation frequencies.

\begin{figure}
\centering
\includegraphics[]{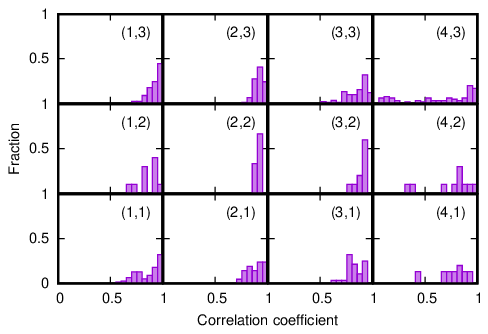}
\caption{Distribution of correlation coefficients for non-rotating models in 
each region.}
\label{modcor}
\end{figure}

\begin{figure}
\centering
\includegraphics[]{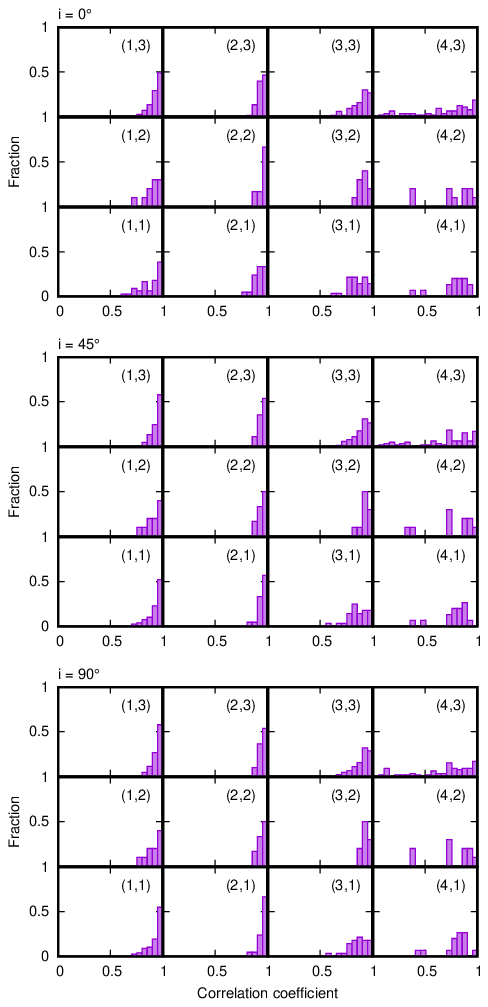}
\caption{Distribution of correlation coefficients when rotational splitting
is applied.  The inclination of the axis of rotation is fixed for all stars. 
The inclinations are $i = 0^\circ$ (top panel) $i = 45^\circ$ (middle panel) 
and $i = 90^\circ$ (bottom panel).}
\label{modrot}
\end{figure}

Choosing a bin size $\delta\nu = 4, 5$ or 6\,d$^{-1}$ results in only 
slight differences in the resulting distribution of correlation coefficients
for the the models and almost no difference at all for the observations.
Below $\delta\nu = 3$\,d$^{-1}$ the distributions begins to show an
increasing noise level which tends to obscure any correlation that may be
present.  This is to be expected as a smaller bin size leads to less overlap 
between distributions.  If the value of $\delta\nu$ is too large, the overlap 
between frequency distributions will increase and the distribution of $r$ will 
flatten and start to loose its discriminatory power.  This begins to occur when 
$\delta\nu = 7$\,d$^{-1}$ in the models. A value of $\delta\nu = 5$\,d$^{-1}$ 
was chosen as being a good compromise.

\section{Correlation from pulsation models}

A small change in effective temperature and luminosity in a model of a 
non-rotating star will lead to small changes in the predicted frequencies.  It 
is therefore expected that the frequency distributions derived from models with
similar values of ($\log T_{\rm eff}$, $\log L/L_\odot$) will be strongly 
correlated.  The aim of this section is to show that this is indeed the case.  
It is also important to study how the correlation coefficient might change 
between models with different rotation rates.  When calculating the
correlation coefficient between model spectra, the amplitudes are taken to
be proportional to the growth rate.  In addition, a visibility factor which
depends on the spherical harmonic degree, $l$, is applied.

For a model in a particular region, the correlation coefficient between
the frequency pattern of unstable modes was correlated with the frequency
pattern for all other models in the region.  The number of models for each
region, $N_{\rm mod}$, is shown in Table\,\ref{regions}.  Each pair of models 
gives rise to one correlation coefficient, $r$.  The resulting distribution of 
$r$ for all pairs of non-rotating models is shown in Fig.\,\ref{modcor}.  As 
expected, the distribution of $r$ tends to concentrate near $r = 1$.

\citet{Reese2006} provides a formula for third-order rotational perturbation 
as well as the necessary coefficients for $l \le 3$.  For higher values of 
$l$, coefficients for $l = 3$ were used.  The distribution of rotational 
equatorial velocities for mid-A stars \citep{Balona2022b} was used to derive 
the distribution of rotational frequencies required in the formula.

To test the effect of rotation, frequency splitting was applied to each 
unstable mode with spherical harmonic degree $l$, so that a frequency in 
the non-rotating model is replaced by $2l+1$ frequencies.  The relative 
amplitudes of the multiplets vary according to $P_l^{|m|}(\cos i)$ (Eq.\,6 of 
\citealt{Dziembowski1977} or Eq.\,17 of \citealt{Gizon2003}).  The resulting
distributions are shown in Fig.\,\ref{modrot} for three fixed inclination
angles of $i = 0^\circ$, $i = 45^\circ$ and $i = 90^\circ$.  Once again, the
values of $r$ tend to be close to unity.

The models indicate that one should expect a high level of correlation 
of frequency patterns for $\delta$~Sct stars within a given region. 
Furthermore, it suggests that differing rotation rates among the stars do
not affect the correlation very much at all.

\section{Correlation from observations}

As can be seen from Fig.\,\ref{reg} and Table\,\ref{regions}, each region is 
well populated with $\delta$~Sct stars.  On average, there are about 600 stars 
in each of the 12 regions. It is important that a sufficiently large number of 
stars is used because the errors in $\log T_{\rm eff}$ and $\log L/L_\odot$ are
comparable to the region size.  Systematic errors are also present.  For 
example, rapid rotation tends to shift the apparent effective temperature due 
to gravity darkening.  The mean equatorial velocities for A/F stars is around 
$v_e \approx 80$\,km\,s$^{-1}$ \citep{Balona2022b}, whereas the typical 
critical rotational velocity is $v_c \approx 400$\,km\,s$^{-1}$ ($v_e/v_c
\approx 0.2$). Significant systematic effects due to gravitational darkening 
only begin for $v_e/v_c \gtrapprox 0.5$ (see Fig.\,4 in \citealt{Salmon2014}), 
so this effect is likely to be small.  

The visibility of a pulsation peak also depends on the angle of inclination, 
$i$.  A variation in $i$ leads only to a variation in amplitude, not frequency,
but will still tend to lower the correlation. Since $i$ is randomly distributed, 
most stars will be observed close to $i \approx 90^\circ$ because the number of 
stars observed with inclination $i$ is distributed as $\sin i$.  For this 
reason, the effect of differing angles of inclination is likely to be small as 
well.

Frequencies of $\delta$~Sct stars were extracted from the periodograms.  The
extracted frequency is deemed significant if the signal-to-noise ratio in the 
amplitude exceeds 4.7.  Each significant extracted frequency peak was 
assigned a weight proportional to its amplitude when calculating the
histogram of the frequency distribution.  As before, the correlation 
coefficients, $r$, were obtained for all possible pairs of stars within a given 
region.  Finally, the distribution of $r$ for each region was determined.

\begin{figure}
\centering
\includegraphics[]{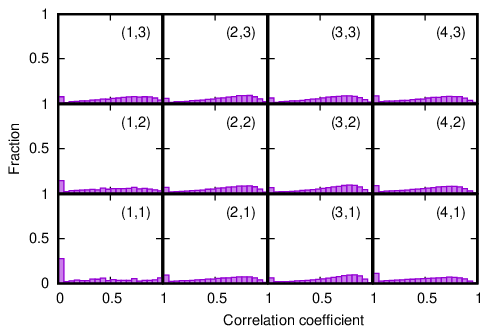}
\caption{Distribution of correlation coefficients for {\em TESS}
$\delta$\,Sct stars within a small region in the H--R diagram.}
\label{obs}
\end{figure}

The calculation was performed for 1673 stars with precise measurements
of effective temperature as well as 7353 stars for which any estimate of
effective temperature was accepted.  The results are very similar.  In
Table\,\ref{regions}, $N_{\rm obs}$ is the number of stars in each region
for the latter case.  The distributions of correlation coefficients are shown 
in Fig.\,\ref{obs}.  

Fig.\,\ref{obs} shows that, contrary to the model predictions, there is very
little correlation between the frequency patterns in $\delta$~Sct stars with 
similar effective temperatures and luminosities.  As discussed above,
rotation does not affect the correlation coefficient very much at all. 
Therefore, one can rule out rotation as the cause of the observed small
correlations.

\section{Mode coupling}

Mode coupling involving non-linear mode interactions between two or more modes 
may result in a complex frequency spectrum which could account for the wide 
disparity in the frequency patterns in $\delta$~Sct stars. 
\citet{Mourabit2023} used the theory of three-mode coupling to study the 
strength and prevalence of non-linear mode interactions in $\delta$~Sct models 
across the instability strip.  They conclude that resonant mode interactions 
can be signiﬁcant.  Mode coupling may explain the rapid changes in amplitude 
and frequency seen in some $\delta$~Sct stars \citep{Bowman2016, Bowman2021b}.

A combination frequency involving two parent modes, $\nu_1, \nu_2$  is given
by $\nu = n_1\nu_1 + n_2\nu_2$ where $n_1$ and $n_2$ are positive or
negative integers.  If the frequency $\nu$ matches an observed frequency peak 
to within the observational error, then $\nu$ can be regarded as a combination 
frequency.  To test the degree of mode coupling that may be present in 
$\delta$~Sct stars, all possible combination frequencies with $|n_1| \le 7$
and $|n_2| \le 7$ were identified among 6840 $\delta$~Sct stars with at
least five significant peaks.  

\begin{figure}
\centering
\includegraphics[]{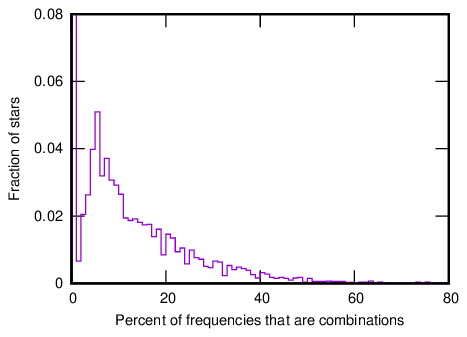}
\caption{Fraction of $\delta$~Sct stars as a function of percentage of
frequencies that are combinations.}
\label{comb}
\end{figure}

It was found that at least one or more combination frequencies are present in 
4151 stars (60\,percent of the total), but in most of these stars only a small
fraction of the frequency peaks are identified as combination frequencies.
The distribution of stars with combination frequencies is shown in 
Fig.\,\ref{comb}. 

Higher order combinations involving three or even more modes could occur,
but these would be difficult to identify.  Considering that very few
combination frequency peaks were identified in most stars, it is very unlikely 
that mode interaction may be the explanation for the diversity of frequency 
patterns.

\section{Period--luminosity relation} 

Considering the large disparity in frequency patterns at constant
temperature and luminosity, it would be surprising if a period--luminosity
law, similar to the Cepheid P--L law, can be found for $\delta$~Sct stars.  
Nevertheless, the existence of a P--L relation in $\delta$~Sct stars has
been reported from time to time.  Since there are usually many periods to be 
chosen in any given $\delta$~Sct star, the period corresponding to the highest 
amplitude is used.  \citet{Barac2022} found that stars lie on a 
ridge that corresponds to pulsation in the fundamental radial mode, while 
others pulsate in shorter period overtones.  The correlation is best seen in 
the period - mean density relationship.

\begin{figure}
\centering
\includegraphics[]{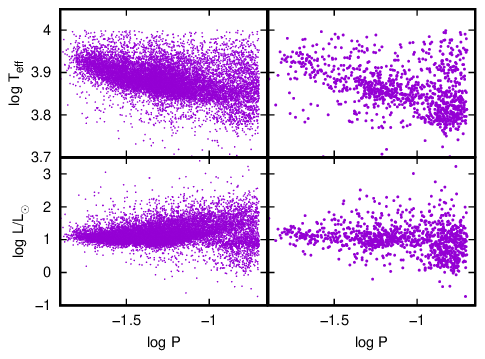}
\caption{Effective temperature and luminosity as a function of pulsation
period, $P$ (days), using the period which corresponds to the peak with highest 
amplitude in {\em TESS} $\delta$~Sct stars.  The left panels include all
stars, while the panels on the right are restricted to stars with one
dominant frequency.}  
\label{pl}
\end{figure}

It should be borne in mind that there are several correlations between
the pulsation period, $P$, and the physical parameters.  Pulsations in 
$\delta$~Sct stars are mainly driven by the $\kappa$ opacity mechanism 
operating in the He\,II partial ionization zone.  In cool stars, the zone is 
deep, the resonant cavity is large and the radial order of the pulsations is 
low, so the periods are typically long.  In hot stars, the zone is shallow, 
the resonant cavity is small and the radial order is high, so the periods are 
shorter.  This effect can be seen in the model frequencies in Fig.\,\ref{freq}.  

From the observations of 11833 $\delta$~Sct stars, the frequency of maximum
amplitude was extracted.  The left panels of Fig.\,\ref{pl} shows the
relationship between $\log T_{\rm eff}$ and $\log P$ (top panel) and between
$\log L/L_\odot$ and $\log P$ (bottom panel).  As predicted by the models,
the pulsation period decreases as the temperature increases, though the
correlation is not particularly strong.  However, there is no significant
correlation between  $\log P$ and $\log L/L_\odot$.

It is possible that stars with a single dominant mode may be best suited to
test the idea of a P--L relationship because these could be radial modes (or
at least non-radial modes of the same degree).  The corresponding
relationships for these stars are shown in the right panels of Fig.\,\ref{pl}.
Once again, the only correlation is between $\log T_{\rm eff}$ and $\log P$.  
This indicates that modes in stars with only one dominant frequency do not 
have the same value of $l$ or radial order.

In conclusion, there seems to be a weak correlation between the
dominant pulsation period and effective temperature in $\delta$~Sct stars,
in agreement with model predictions.  \citet{GarciaHernandez2015} analyzed
$\delta$~Sct stars in eclipsing binaries and derived an observational scaling 
relation between the stellar mean density and a frequency pattern in the 
oscillation spectrum which is independent of rotation rate.  It is possible
that careful analyses of selected stars with well-determined physical
parameters in this manner might lead to reliable mode identification in some 
$\delta$~Sct stars.  However, a universal period -- luminosity law similar to 
that in Cepheids does not seem to exist.

\section{Discussion and Conclusions}

The surprisingly wide variety of frequency patterns in $\delta$~Sct stars has 
been noted several times (e.g. \citealt{Balona2015d, Bedding2023}).  Inspection
of many thousands of periodograms gives the impression that each $\delta$~Sct 
star has a unique set of frequencies, unrelated to its location in the H--R 
diagram.  

In this paper, periodograms of $\delta$~Sct stars observed by {\em TESS} are 
presented.   These are shown for stars that lie within small regions of the
H--R diagram.  In each region, where stars have essentially the same effective
temperature and luminosity,  stars can be found with just a single frequency
peak and also with many frequency peaks.  This is in contradiction to 
pulsation models which predict that stars with the same global parameters 
and rotation rates should pulsate with the same set of frequencies. 

In order to obtain an objective measure of similarity of frequency patterns
in a pair of stars, a correlation method was developed. Using non-adiabatic 
non-rotating pulsation models, it is found that the frequencies of unstable 
modes within a  small effective temperature and luminosity range are indeed 
correlated.  The effect of rotation was simulated by applying rotational 
splitting using a 3rd-order approximation formula \citep{Reese2006}.  A spread 
of rotation periods similar to that expected for mid-A stars was used.  The 
results show that the correlations among rotating stars are as high as among 
non-rotating stars.

When applied to a large sample of {\em TESS} $\delta$~Sct stars with similar
effective temperatures and luminosities, very little correlation is
found.  In other words, the observed frequencies are not related to the 
location of the star in the H--R diagram at all.  This is in contradiction
to what is found in the models.

The possibility that combination frequencies may be responsible for some of 
the discrepancy seems unlikely.  Although combination frequencies were detected
in about 60\,percent of the stars, they constitute only a small fraction of
the total number of frequency peaks observed in a star.

The fact that frequencies in $\delta$~Sct stars with similar effective
temperatures and luminosities vary so widely indicates that there is no
possibility of finding a useful period - luminosity relation.  Even for
stars where only one dominant mode is present, the frequency of this mode
can vary widely between stars with essentially the same physical parameters. 
This means that the spherical harmonic degree and/or radial order must
differ from star to star.  However, there is a general trend for the hotter, 
more luminous stars to have shorter pulsation periods, in agreement with 
non-adiabatic pulsation models.

The effect described here suggests that there is a fundamental problem with 
current models. It is clear that the excitation of modes is a highly
non-linear process which perhaps involves mode interaction.  Current models
also do not accurately describe the outer layers of A and F stars.  The likely 
presence of starspots and even flares on early-type stars \citep{Balona2021b} 
indicates a complex structure. It is possible that local conditions may
determine the frequencies that are selected.

Progress might be achieved by including the effect of surface convection, even 
in the hottest models of $\delta$~Sct stars.  An attempt should be made on 
modifying the pulsational parameters in these models to determine if a wide 
variety of unstable frequencies can thereby be obtained.  The possible
effect of photospheric structures on the number of observed pulsation peaks
could be studied using high-dispersion spectroscopy.  Observations of stars
with a single dominant mode and stars with very rich frequency spectra may
provide clues to the difference in mode selection and the extreme sensitivity 
of pulsational driving to local conditions.

\section*{Acknowledgments}

I thank the National Research Foundation of South Africa for financial
support and Dr. W. Dziembowski for permission to use his code.

This paper includes data collected by the {\it TESS} mission. Funding for the 
{\it TESS} mission is provided by the NASA Explorer Program. Funding for the 
{\it TESS} Asteroseismic Science Operations Centre is provided by the Danish 
National Research Foundation (Grant agreement no.: DNRF106), ESA PRODEX
(PEA 4000119301) and Stellar Astrophysics Centre (SAC) at Aarhus University. 
We thank the {\it TESS} and TASC/TASOC teams for their support of the present
work.

This work has made use of data from the European Space Agency (ESA) mission 
Gaia (\url{https://www.cosmos.esa.int/gaia}), processed by the Gaia Data 
Processing and Analysis Consortium (DPAC,\\
\url{https://www.cosmos.esa.int/web/gaia/dpac/consortium}).\\ 
Funding for the DPAC has been provided by national institutions, in particular 
the  institutions participating in the Gaia Multilateral Agreement.  

This research has made use of the SIMBAD database, operated at CDS, 
Strasbourg, France.  This research has made use of the VizieR catalogue access 
tool, CDS, Strasbourg, France (DOI: 10.26093/cds/vizier). The original 
description of the VizieR service was published in A\&AS 143, 23.

The data presented in this paper were obtained from the Mikulski Archive for 
Space Telescopes (MAST).  STScI is operated by the Association of Universities
for Research in Astronomy, Inc., under NASA contract NAS5-2655.

\section*{Data availability}

The data underlying this article can be obtained from
{\tt https://sites.google.com/view/tessvariables/home}.

\bibliographystyle{plainnat}
\bibliography{mosel}

\end{document}